\RequirePackage{ifpdf}
\documentclass[12pt]{JHEP3}
\pdfoutput = 1
\usepackage{cite}

\let\P=\Pi

\usepackage{color}
\usepackage[makeroom]{cancel}
\usepackage{multicol}
\usepackage[normalem]{ulem}
\usepackage{epsfig,bm,graphicx,caption}
\epsfclipon
\usepackage{amsmath,amssymb,amsbsy,amstext,amsthm}
\usepackage{subfig}
\usepackage{cancel}
\let\normalcolor\relax  

\newcommand{\roughly}[1]{\mathrel{\raise.3ex\hbox{$#1$\kern-0.85em
\lower1ex\hbox{$\sim$}}}}

\def\be{\begin{equation}}
\def\ee{\end{equation}}
\def\ba{\begin{eqnarray}}
\def\ea{\end{eqnarray}}

\theoremstyle{definition}

\title{Cuscuton Bounce}
\author{Supranta S. Boruah${}^{1}$, Hyung J. Kim${}^{1}$, Michael Rouben${}^{3}$, Ghazal Geshnizjani${}^{1, 2}$\\
$^1$Department of Applied Mathematics, University of Waterloo,  200 University Ave W, Waterloo, Ontario, N2L 3G1, Canada\\ $^2$Perimeter Institute for Theoretical Physics, 31 Caroline St. N., Waterloo, ON, N2L 2Y5, Canada\\$^3$ MBD Consulting Canada Inc., 3 Valloncliffe Rd, Toronto, ON L3T 2W6, CA}

\date{}

\abstract{In general relativity producing a regular bounce entails violation of Null Energy Condition for a dynamical source in the model. That generically indicates existence of ghosts or other instabilities. However, in cuscuton modification of gravity, the correspondence between a background bounce and violation of Null Energy Condition for dynamical sources is broken. Cuscuton action, modifies equations of motion in Infra Red limit allowing the background to go through a regular bounce phase. At the same time, since it does not contain any dynamical degrees of freedom, it does not lead to ghosts or other instabilities. Here, we present a toy scenario of a regular bouncing cosmology and prove this claim. Our model is presented as a proof of concept at this point and does not aim to explain observations in late time cosmology. }
\preprint{}


\begin{document}

\section{Introduction}
The last decade or two has been called the age of precision cosmology. Precise observations of the Cosmic Microwave Background (CMB) radiation and Large Scale Structures (LSS) have provided tight constraints on our cosmological models. Cosmology on theoretical front has also been very successful in building models of early universe that can match these observations. The inflationary paradigm is arguably the most popular among the current models. However, inflationary models do not address all the fundamental questions about the beginning of universe. For instance, it has been argued that inflationary space-times are not past-complete\cite{BGVThm}. In other words, inflation doesn't provide a resolution to singularity problem. It is generally posited that quantum gravity effects might lead to the resolution of this problem. However, invoking the unknown powers of quantum gravity to address any initial condition problem that we can not resolve, can be a double-edged sword. For instance, if quantum gravity effects are important, the framework of quantum field theory on curved space-time, which is used to make predictions for inflation becomes invalid at the scales of interest and leads to the so-called trans-Planckian problem \cite{Trans_Planckian, Brandenberger_review}. One would hope that if quantum gravity is relevant at early universe, its effects can be formulated in systematic ways, that can also be tested.

One way in which singularity problem can be evaded is by considering regular bouncing cosmologies, where an initially contracting universe, `bounces' and starts expanding. Many models of regular bouncing cosmologies have been proposed in the literature \cite{Ekpyrotic, Ijjas_Steinhardt, Angelika, Cai_Brandenburger, Cai_Brandenburger2, GBounce,GalileanBounce,GhostCondensate,Cai:2016thi,Cai:2017tku, DubovskyNEC,Gal_genesis, Dobre:2017pnt, Ijjas:2016tpn,Creminelli:2016zwa}. Many of these models share a common feature of possessing scalar field components since scalars provide the simplest framework to describe dynamics. However, the actions for scalar fields differ from each other, depending on which fundamental conjectures they are inspired from. These conjectures can be motivated from phenomenological theories of modified gravity, string theory, loop quantum gravity, etc. One of the obstacles that these scenarios face is that in general relativity, a regular bounce requires the violation of Null Energy Condition (NEC). This generically leads to instabilities or superluminal speed of sound\footnote{We refer readers to \cite{DubovskyNEC,Sawicki:2012pz} for further reading and to \cite{Rubakov:2014jja} for a good review on how NEC violation can lead to instabilities, superluminality or possibly unbounded Hamiltonians from below.}. There are few proposals in the literature regarding stable ways to violate the NEC. Ghost condensate\cite{GhostCondensate} is one of the early models that was suggested to produce a healthy regular bounce. Later, it was noticed that in the context of late universe cosmology \cite{kineticBraidingIntro}, a sub-class of Horndeski actions, `Kinetic Gravity Braiding', can have an healthy \cancel{NEC}. This led to the development of more improved versions of the regular bouncing scenarios \cite{GBounce, GalileanBounce} within Horndeski theories. The stability and superluminal nature of these models has been a subject of interesting debates in literature. Authors in \cite{Creminelli:2012my,Ijjas:2016tpn,Ijjas_Steinhardt} have argued that it is possible to obtain a healthy bounce using Galilean action while \cite{Easson:2013bda,Dobre:2017pnt} argue that when coupling to matter or other regions of phase-space are included, Galilean models have superluminal speed of sound\footnote{Since the literature on this topic is very extensive and still developing, we refer readers to references and citations of the mentioned papers for further details.}. 

It is also worth noting that interest in healthy \cancel{NEC} also extends to other areas of gravitational physics, such as traversable wormhole solutions, or models which require universe to be initially static\cite{Gal_genesis}. In the case of traversable wormhole, it has been shown that there are some no-go theorems that apply\cite{Rubakov:2016zah,Rubakov:2015gza}. 


In this paper, we present a new resolution for instabilities associated with \cancel{NEC} scenarios. 
We show that cuscuton modification of gravity\cite{Cuscuton_original, Cuscuton_cosmology} allows for an effective violation of NEC in FRW backgrounds while the actual matter sources satisfy NEC. Note that cuscuton field, mimics the appearance of adding a non-canonical scalar field (cuscuton field) to general relativity. However, the kinetic term of this field is such that it has no dynamical degree of freedom\footnote{``No dynamical degree of freedom" can be interpreted as,  equation of motion for cuscuton field does not have any time derivatives. This can be shown explicitly at linear order in flat space-time or around FRW backgrounds. Since action is covariant, that implies there are no local degrees of freedom. For more details, we refer readers to \cite{Cuscuton_paper}.  In \cite{Daniel_Hamiltonian}, authors argue that in Hamiltonian formalism Cuscuton acts as a dynamical field when its inhomogeneities are considered. We suspect that the corresponding equations are not well-posed and the duality of Hamiltonian formalism to Lagrangian formalism is breaking down for Cuscuton. That is an interesting topic that requires further investigation.} but it modifies gravity in Infrared (IR) regime. Due to its non-dynamical nature, cuscuton models still need other fields to produce dynamics. In other words, cuscuton is instrumental to make the background bounce but the actual dynamical degree of freedom does not violate NEC. Therefore, our model does not fall under the single field  $P(X,\phi)$ models that violate NEC and the problems discussed in  \cite{Vikman:2004dc, Easson:2016klq, ClaudiaUnitarity} do not apply to our model. 

We would like to also point out that cuscuton terms have previously been shown to be important in having consistent background condition for generating a bounce solution within k-essence models \cite{Romano_Enea} \footnote{There, cuscuton term is part of the single field non-canonical kinetic terms  and instabilities discussed in \cite{Vikman:2004dc, Easson:2016klq, ClaudiaUnitarity} can be applicable.} as well as a stable matter bounce scenario in massive gravity models \cite{Lin:2017fec}. 

Our paper is structured in the following way. In Section \ref{bounce}, we present a toy model for a cuscuton bounce scenario. In Section \ref{perturbs}, we analyze the existence of ghosts and other instabilities in this model. We end with our concluding remarks in Section \ref{conclusion}.

\section{A toy model for cuscuton bounce} \label{bounce}

Consider the following action for a scalar field with a noncanonical kinetic term,
\begin{eqnarray}
S=\int d^4x \sqrt{-g} L(\varphi, X)~,\label{gaction}
\end{eqnarray}
where $L$ is an arbitrary function of the scalar $\varphi$ and
$X\equiv\frac{1}{2}\partial_\mu\varphi\partial^\mu\varphi$ \footnote{We will use
units with $M_p^2=1/8\pi G$ and the metric signature is
$(+,-,-,-)$.}.

This action is compatible with a perfect fluid description
\be
T_{\mu\nu}=( \rho+P) u_{\mu}u_{\nu}-P g_{\mu\nu}
\ee
assuming 
\ba
u_{\mu} & \equiv & \frac{\partial_\mu \varphi}{\sqrt{2X}}
\ea
is time-like. The energy density and the pressure in the comoving fame of $u_\mu$ are  
\begin{eqnarray}
\rho&=&T_{\mu\nu}u^\mu u^\nu =2XL_{,X}-L\\P&=&L.
\end{eqnarray}
We use  $_{,X}$ to denote the partial derivative with respect to the variable
$X$. 

In a flat FRW background 
\begin{eqnarray}
ds^2=dt^2-a^2(t)\delta_{ij}dx^idx^j,
\end{eqnarray}
the homogeneous field equation (\ref{gaction}) reduces to,
\begin{eqnarray} \label{EQMcus}
(L_{,X}+2XL_{,XX})\ddot\varphi_0+3HL_{,X}\dot\varphi_0+L_{,X\varphi}\dot\varphi^2_0-L_{,\varphi}=0~,
\end{eqnarray}
where $H$ represents Hubble constant and we denote the time derivative with an overdot. {\it Cuscuton} modification of gravity is achieved by taking the in-compressible limit of the above perfect fluid such that everywhere on $(\varphi, X)$ plane
\be
L_{,X}+2XL_{,XX}=0. 
\ee
 As we see in that limit, the equation of motion is no longer second order since the second time derivative of $\varphi$ vanishes (see \cite{Cuscuton_original, Cuscuton_cosmology} for more details). 
A Lagrangian that satisfies the above requirement everywhere in phase space corresponds to  
\begin{eqnarray}
L(\varphi, X)=\pm \mu^2\sqrt{2X}-V(\varphi), 
\end{eqnarray}
which is called {\it Cuscuton} Lagrangian  \footnote{It had been already noted in \cite{Cuscuton_origin_vikman} that this Lagrangian corresponds to the $c_s \rightarrow \infty$ limit in a $P(X, \phi)$ theory.}. What is more interesting about this Lagrangian is that when we substitute it in \ref{EQMcus}, not only $\ddot{\varphi}$ dependence vanishes but that $\dot{\varphi}$ dependence cancels as well, leading to the following constraint equation,
\begin{eqnarray}\label{cus_H2}
\pm \textrm{sign}(\dot\varphi)3\mu^2H+V^{\prime}(\varphi)=0~. 
\end{eqnarray}
$\mu$ can in principle depend on $\varphi$ but that dependence can be absorbed into a field redefinition such that a new cuscuton action with constant $\mu$ and a new potential is obtained. 

Since the cuscuton equation is not dynamical, contributions of dynamical matter sources in the universe are necessary to obtain any cosmological evolution. Here, we consider a toy bounce model where the universe consist of a barotropic component $p_m=w \rho_m$ in addition to cuscuton field. A desirable model would initially be a contacting universe where in very early times the cuscuton modifications of gravity are negligible. However, as it gets smaller the cuscuton modification becomes important, causing the universe to bounce into an expanding phase. For simplicity we assume $w=1$ so $\rho_m\propto a^{-6}$, making cuscuton contributions grow even faster close to the bounce and be dominant over anisotropies. However, this assumption is not fundamental for our result. A simple way to produce such an equation of state from action is to include a minimally canonical scalar field, $\pi$, with no potential. That will later allow us to  consistently study the behaviour of perturbations during the bounce. 

A main feature of a regular bounce ($H\neq \pm\infty$) is that universe goes from a contracting phase ($H<0$) into an expanding one ($H>0$) at finite value of scalar factor, $a_b$. 
This criteria automatically implies
\ba
H_b&=&0 \label{H0}\\
 \dot{H}_b&>&0, \label{hdotpositive}\
\ea
where $b$ denotes the bounce. In general relativity, the second condition necessitates the violation of NEC for a perfect fluid source. 

We now investigate the possibility of a bounce solution in a framework, consisting of cuscuton and a barotropic matter source $\rho_m$.  

The Friedmann and continuity equations can be obtained from action or Einstein's equations
\begin{eqnarray}
H^2&=&\frac{1}{3M_p^2}[ V(\varphi)+\rho_m] \label{FRDeq}\\
\dot{H}& = &-\frac{1}{2M_p^2}[\pm \mu^2\sqrt{2X}+(1+w)\rho_m]. 
\end{eqnarray}
Therefore, requiring the energy condition $\rho_m>0$ and \eqref{H0} be satisfied at the bounce leads to
\be
V(\varphi_b)<0.
\ee 
On the other hand, condition \eqref{hdotpositive} implies that only the choice of the negative sign for cuscuton kinetic term could lead to a bounce solution. So from here on we only consider 
\begin{eqnarray} \label{Lagrangian}
L(\varphi, X)=- \mu^2\sqrt{2X}-V(\varphi).
\end{eqnarray}

This in turns, yields Eq. (\ref{cus_H}) becomes
\begin{eqnarray}\label{cus_H}
- \textrm{sign}(\dot\varphi)3\mu^2H+V^{\prime}(\varphi)=0~, 
\end{eqnarray}
which leads to   
\be \label{signVdd}
3\mu^2\dot{H}=V^{\prime\prime}(\varphi)|\dot{\varphi}|.
\ee
Therefore, in the regimes that $NEC$ is valid ($\dot{H}<0$), cuscuton potential must satisfy  
\be 
V^{\prime\prime}<0, ~~~~ \text{for    NEC}, \label{V''NEC}
\ee 
but close to the bounce, 
\be 
V^{\prime\prime}>0,  ~~~~\text{for    \cancel{NEC}}. \label{V''bounce}
\ee
In addition, substituting $H$ from Eq. (\ref{cus_H2}) back into Eq. (\ref{FRDeq}), we arrive at 
\begin{eqnarray}\label{cus_sad}
\frac{M_p^2}{3\mu^4}V^{\prime 2}(\varphi)=V(\varphi)+\rho_m~.
\end{eqnarray}
This equation demonstrates how for a particular potential $V(\phi)$, the evolution of cuscuton depends on other matter sources in the universe. We can also use this relation to derive further constraints on cuscuton potential.
Taking a time derivative of Eq. (\ref{cus_sad}), combining it with continuity equation for matter source, $\dot{\rho}_m=-3H(\rho_m+p_m)$, and Eq. (\ref{cus_H2}) we get\footnote{This equation together with \eqref{cus_sad} also demonstrates, how $\dot{\varphi}$ is uniquely determined as a function of $\varphi$.}
\ba \label{Vsecondd}
\frac{2 M_p^2}{3\mu^4}V^{\prime \prime}(\varphi)- 1=-(1+w){\frac{\rho_m}{\mu^2 |\dot{\varphi}|}}<0, 
\ea
for $w>-1$ or that
\ba
V^{\prime \prime}(\varphi)< \frac{3\mu^4}{2 M_p^2}.
\ea
This enables us to conclude that while the shape of the potential in the \cancel{NEC} era (around the bounce) is convex (\ref{V''bounce}), its convexity is in this range:  
\be \label{ddVbounce}
0<V^{\prime \prime}(\varphi_b)< \frac{3\mu^4}{2 M_p^2}.
\ee
However, as we argued before, potential has to become concave, $V^{\prime \prime}(\varphi)<0$, in regions that NEC is restored. Setting additional assumptions, such as when far from the bounce, cuscuton modifications of gravity are negligible, can also be used to obtain additional restriction about the shape of the potential. This assumption can be applied by requiring  
\ba 
&&\lim_{t \to \pm \infty} H^2  =\lim_{t \to \pm \infty}  {\frac{1}{3 M_p^2}}\rho_m \to 0, \\
&&\lim_{t \to \pm \infty} \dot{H}  =\lim_{t \to \pm \infty}  -\frac{1+w}{2M_p^2}\rho_m\to 0, 
\ea 
where $t=0$ corresponds to the bounce. Using the above conditions in combination with Eq. (\ref{Vsecondd}), Eq. (\ref{cus_H2}) and Eq. (\ref{cus_sad}), one can show  
\ba
&&\lim_{t \to \pm \infty} V^{\prime}(\varphi_{\infty})=0 \label{vinfty}\\
&&\lim_{t \to \pm \infty}  \frac{3\mu^4}{M_p^2}\frac{|V(\varphi_{\infty})|}{V^{\prime 2}(\varphi_{\infty})}\ll 1 \label{dvinfty}\\
&&\lim_{t \to \pm \infty} V^{\prime\prime}(\varphi_{\infty})\ll - \frac{3\mu^4}{\sqrt{2}M_p^2} \label{ddVinfty}
\ea
Note that Eq. \eqref{ddVbounce} and Eq. \eqref{ddVinfty} imply
\be
|V''(\pm\varphi_{\infty})|/ V''(\varphi_b)\gg 1. 
\ee
We now introduce a toy model, where the potential contains a quadratic, an exponential and a constant term, such that it meets all the above conditions:

\be \label{potential} 
V(\varphi)\equiv m^2(\varphi^2-\varphi_\infty^2) - {m^4} [e^{(\varphi^2-\varphi_\infty^2)/m^2} -1], 
\ee
with
\ba
\frac{\varphi_\infty^2}{m^2} &\gg& 1.
\ea

The constant term is set to a value that ensures $V(\varphi)=0$ at $\varphi=\pm \varphi_\infty$ and the large value of ${\varphi_\infty^2/ m^2} $, guarantees that $|V''(\pm\varphi_{\infty})|/ V''(\varphi_b)\gg 1$. The viable range for $\mu$ consistent with Eq. (\ref{ddVbounce}) and Eq. (\ref{ddVinfty}) is then
\be
\frac{m^2}{M_p^2}\ll \frac{\mu^4}{M_p^4}\ll \frac{\varphi_\infty^2}{M_p^2}. 
\ee

\begin{figure}[htbp]
\includegraphics[width=\linewidth]{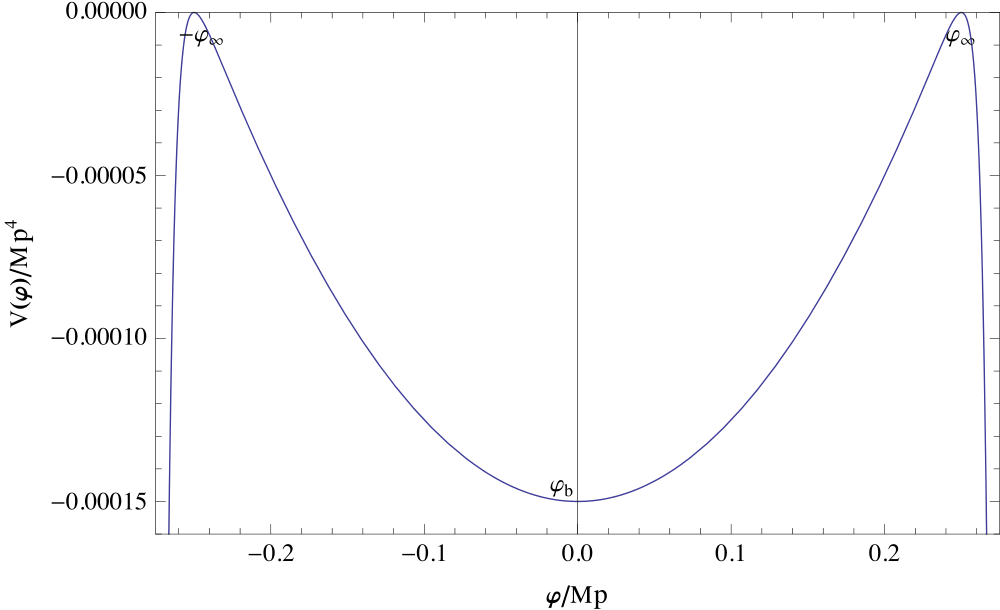}
\caption{$V(\varphi)$ as a function of  $\varphi$ for $m=0.05 M_p$, ${\varphi_\infty^2/ m^2}=25$ .} \label{fig:varphi}
\end{figure}

\begin{figure}[htbp]
\includegraphics[width=\linewidth]{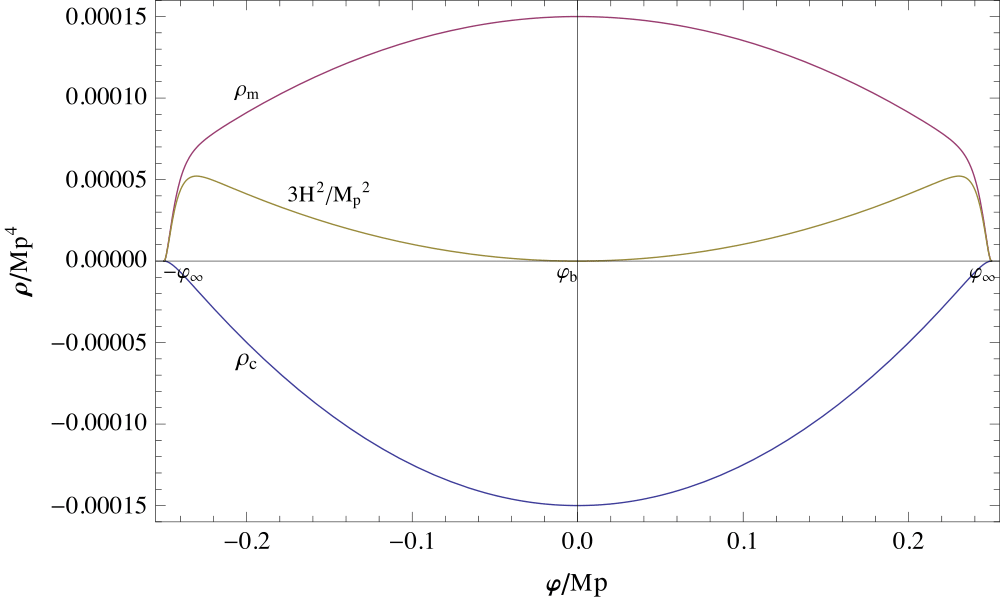}
\caption{Densities and Hubble as functions of  $\varphi$ for $m=0.05 M_p$, ${\varphi_\infty^2/ m^2}=25$ and $\mu=0.3 M_p$.} \label{fig:densityphi}
\end{figure}

\begin{figure}[htbp]
\includegraphics[width=\linewidth]{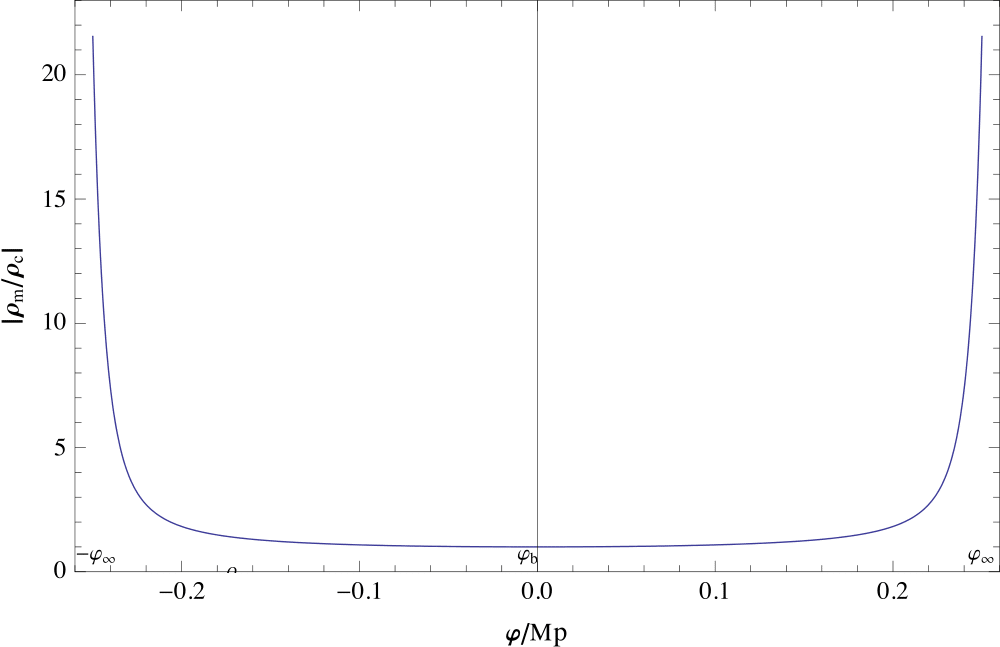}
\caption{Ratio of densities as functions of  $\varphi$ for $m=0.05 M_p$, ${\varphi_\infty^2/ m^2}=25$ and $\mu=0.3 M_p$. For this choice for the values of the parameters in the model, $\rho_{cus}$ becomes more than twenty times smaller than $\rho_m$ far away from the bounce.} \label{fig:densityphirat}
\end{figure}

Figure (\ref{fig:varphi}) displays a schematic shape of a potential where parameters, $m$ and $\varphi_{\infty}$ are set to $m=0.05 M_p$ and $\varphi_{\infty}=0.25M_p$ so $\varphi_{\infty}^2/ m^2=25$.  
For these choice of parameters the allowed range of $\mu$ is $0.22< \frac{\mu}{M_p}< 0.5$. 
For the rest of the discussion we keep the values of the parameters in our model to be fixed at $m=0.05 M_p$, ${\varphi_\infty^2/m^2}=25$ and $\mu=0.3 M_p$.
Substituting the potential described by Eq. (\ref{potential}) into Eq. (\ref{cus_sad}), one can
derive the evolution of $\rho_m$ and $H$ as functions of $\varphi$. Figure \eqref{fig:densityphi} demonstrates the $\varphi$ dependence of these quantities, including $\rho_{cus}\equiv V(\varphi)$ and figure \eqref{fig:densityphirat} shows the ratio of $\rho_m/\rho_{cus}$.  As expected the magnitude of $\rho_{cus}$ is negligible far from the bounce and it is always less than $\rho_m$, except at the bounce where they cancels off in order to yield $\dot{H_b}=0$. 
 

\begin{figure}[htbp]
\includegraphics[width=\linewidth]{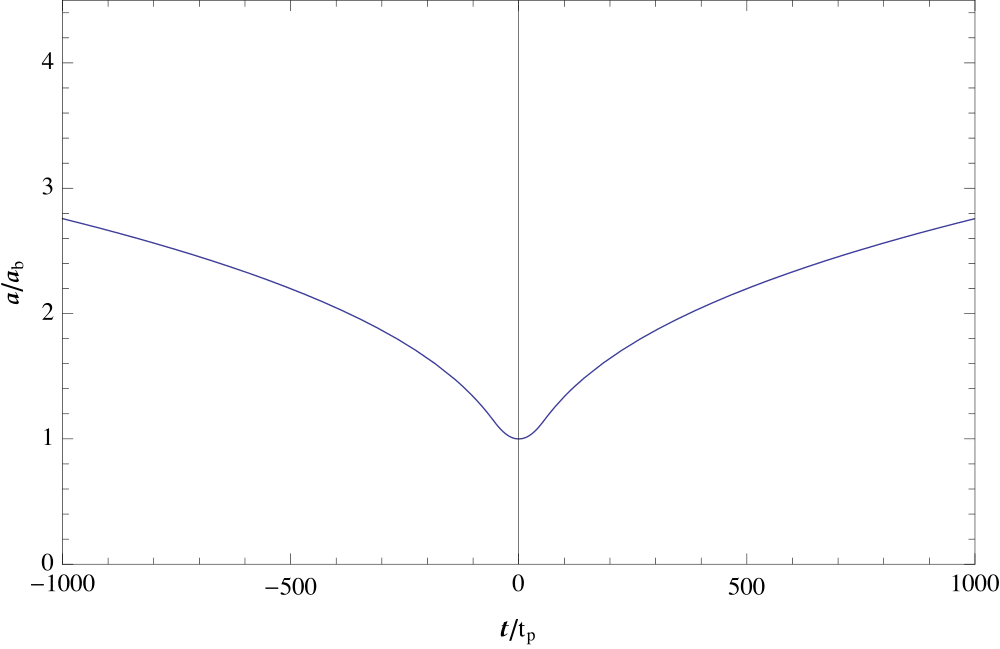}
\caption{The evolution of scale factor, $a(t)$ in time is consistent with universe contracting, undergoing a regular bounce and then expanding.} \label{at} 
\end{figure}

\begin{figure}[htpb]
\includegraphics[width=\columnwidth]{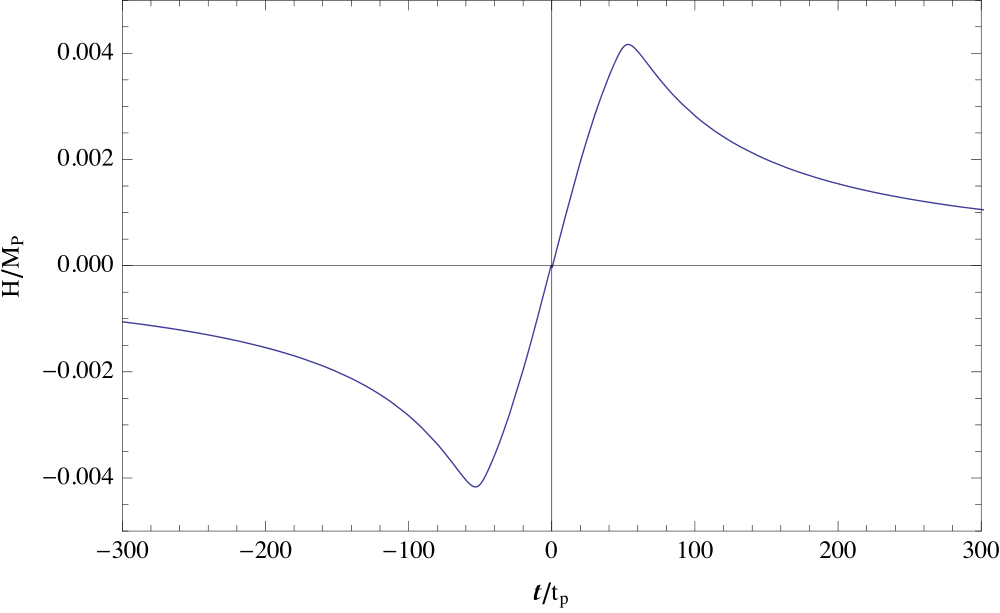}
\caption{The evolution of Hubble constant, $H$, as a function of time. Hubble constant vanishes at the bounce and far from the bounce and there exists a {\it NEC} violating region around the bounce where $\dot{H}>0$.}\label{HT}
\end{figure}

Assuming $\omega=1$ ($\rho_m(\varphi)=\rho_b (a/a_b)^{-6}$), one can obtain the evolution of background parameters numerically in terms of cosmic time or conformal time. Figures (\ref{at}) and (\ref{HT}) illustrate that the cosmological evolution of Scale factor, $a(t)$, and Hubble constant, $H(t)$, are consistent with our picture for a regular bounce cosmology.
Note that for simplicity we have chosen $sign(\dot\varphi)>0$ so $\varphi<0$ coincides with 
Hubble parameter being negative and universe
contracting. Therefore, when $\varphi$ evolves into the positive region, the
universe undergoes a smooth bounce and enters an expanding
phase.


Having developed a consistent picture of the background bounce, next we study the behaviour of cosmological perturbations around this background.

\section{Perturbations in cuscuton bounce}\label{perturbs}

\subsection{Absence of ghosts in cuscuton bounce}
 
One of the generic instabilities that occurs in \cancel{NEC} models is ghost instability. 
This instability is by definition, a UV instability which can be identified though a wrong sign of kinetic term for excitations around flat space-time.  In order to investigate the existence of such an instability in our cuscuton model, we have to study the corresponding action for quantum fluctuations. 
As we mentioned before, adding a canonical scalar field, $\pi$, that doesn't have a potential to cuscuton action, can automatically produce a dynamical source with $\omega=1$. This allows us to study fluctuation in a framework consistent with the background evolution described in section \ref{bounce}. 

The full action after including the canonical scalar field is given by, 

\begin{equation}\label{action}
    S = \int d^4x \sqrt{-g} \bigg[ \frac{M_p^2}{2}R - \frac{1}{2} D_\mu\pi D^\mu \pi - \mu^2\sqrt{-D_\mu \varphi D^\mu \varphi}-V(\varphi) \bigg],
\end{equation}

where $D_{\mu}$ denotes Covariant derivatives, $\varphi$ represents the cuscuton field and $\pi$ stands for the canonical scalar field. 

This action is in fact a subclass of actions that we have studied in \cite{Cuscuton_paper}. There, we probed the existence of ghosts in cuscuton gravity with a generic canonical scalar field content. We found that in general such models do not contain ghosts. We provide a brief summary of our derivation here as well.  

The framework involves the standard way of perturbing action around a flat FRW spacetime. We then used the Unitary gauge fixing\footnote{The definition of the Unitary Gauge is not unique when we have more than one field. Here we call the gauge where $\delta \pi = 0$ as the `unitary' gauge, as $\pi$ field is the only dynamical field in our theory. Using this point of view, cuscuton can be considered as a non-trivial modification of gravity rather than as an additional field.}, where time slices are taken such that $\pi$ field is the clock and the off-diagonal components of the spatial metric is set to zero. Naively, one would expect two independent scalar degrees of freedom arising from the canonical scalar field and the cuscuton field. However, owing to the non-dynamical nature of cuscuton, we are left with only one independent degree of freedom. We expressed this degree of freedom in terms of $\zeta$, corresponding to curvature perturbations in this gauge. In other words, when expressing the metric in the ADM variables, 

\begin{equation}\label{ADM_metric}
	ds^2 = N^2dt^2-h_{ij}(dx^i+N^idt)(dx^j+N^jdt),
\end{equation}

and setting the off-diagonal components of the spatial metric to zero, $\zeta$ is defined through $h_{ij} = a^2\delta_{ij}(1+2\zeta)$. 

Using the Hamiltonian and the momentum constraints, the lapse and the shift can be expressed in terms of $\zeta$ and its time derivative. In this gauge, the cuscuton equation turns out to be a constraint equation. That equation can be inverted into the Fourier space to obtain a closed form for $\delta\varphi_k$ in terms of $\zeta_k$ and its time derivative as,

\begin{equation}\label{CuscEqnFourier}
    \delta \varphi_k = \dot{\varphi}_{0}\frac{(k/a)^2 H \zeta_k+P \dot{\zeta}_k}{\big[(k/a)^2H^2+(3H^2+P+\dot{H})P \big]},
\end{equation}
where, $ P = \frac{1}{2M_p^2}\dot{\pi}^2_0 $. 
Substituting for all the variables in terms of $\zeta$, back in the action, we obtain the quadratics action to be, 

\begin{equation}\label{2ndorderaction}
    S^{(2)} = \frac{M_p^2}{2}\int d^3k\; dt ~ a z^2 \bigg[ \dot{\zeta}^2_k-\frac{c_s^2~k^2}{a^2} \zeta^2_k\bigg]. 
\end{equation}
 $z(k,t)$ and $c_s(k, t)$ are functions that depend on both time and scale and are given by 
\begin{align}
	c^2_s &\equiv \frac{(k/a)^4 H^2+(k/a)^2\mathcal{B}_1+\mathcal{B}_2}{(k/a)^4 H^2+(k/a)^2\mathcal{A}_1+\mathcal{A}_2} \label{cs} \\
    z^2 &\equiv 2\, a^2 P \bigg(\frac{(k/a)^2+3P}{(k/a)^2 H^2+(P)(3H^2+P+\dot{H})}\bigg). \label{z2}
\end{align}
Here, we have introduced the following notation to simplify the relations 

\begin{align}
	\mathcal{A}_1 &= P(6H^2+\dot{H}+P)\\
    \mathcal{A}_2 &= 3P^2(3H^2+\dot{H}+P)\\
    \mathcal{B}_1 &= P(12H^2+3\dot{H}+P)+\dot{H}(2\dot{H}+\frac{H\ddot{H}}{\dot{H}})\\
    \mathcal{B}_2 &=P^2(15H^2-P+\dot{H})-P\dot{H}(12H^2-2\dot{H}+\frac{3H\ddot{H}}{\dot{H}}) 
\end{align}

As is seen from the quadratic action, \eqref{2ndorderaction}, Cuscuton gravity is free from ghost if the coefficient of the kinetics term, $z^2$ is positive. The terms, $(k/a)^2$ and $P$, appearing in the numerator of $z^2$ are both positive. Hence, positivity of $z^2$ depends on the sign of the denominator. The denominator can be simplified using the background equation to,

\begin{equation}\label{ghost_free}
    (k/a)^2 H^2+P(3H^2+\frac{\mu^2}{2M_p^2}|\dot{\varphi}_0|)    
\end{equation}

Written in this form, it is apparent that the denominator is always positive. Hence, this class of Cuscuton Gravity, including our bounce model is ghost-free\footnote{As discussed in \cite{Cuscuton_paper} the other class with $+\mu^2$ in the Lagrangian, also turns out to be ghost free.}. Furthermore, positivity of denominator and non-vanishing contribution from cuscuton modification, guarantees the absence of any poles in this coefficient regardless of wavelength and at the bounce ($H=0$). 

\subsection{Absence of dynamical instabilities in cuscuton bounce}

We next investigate the dynamical stability of the perturbations in different regimes. 

\begin{figure}[htbp]
        \includegraphics[width=\linewidth]{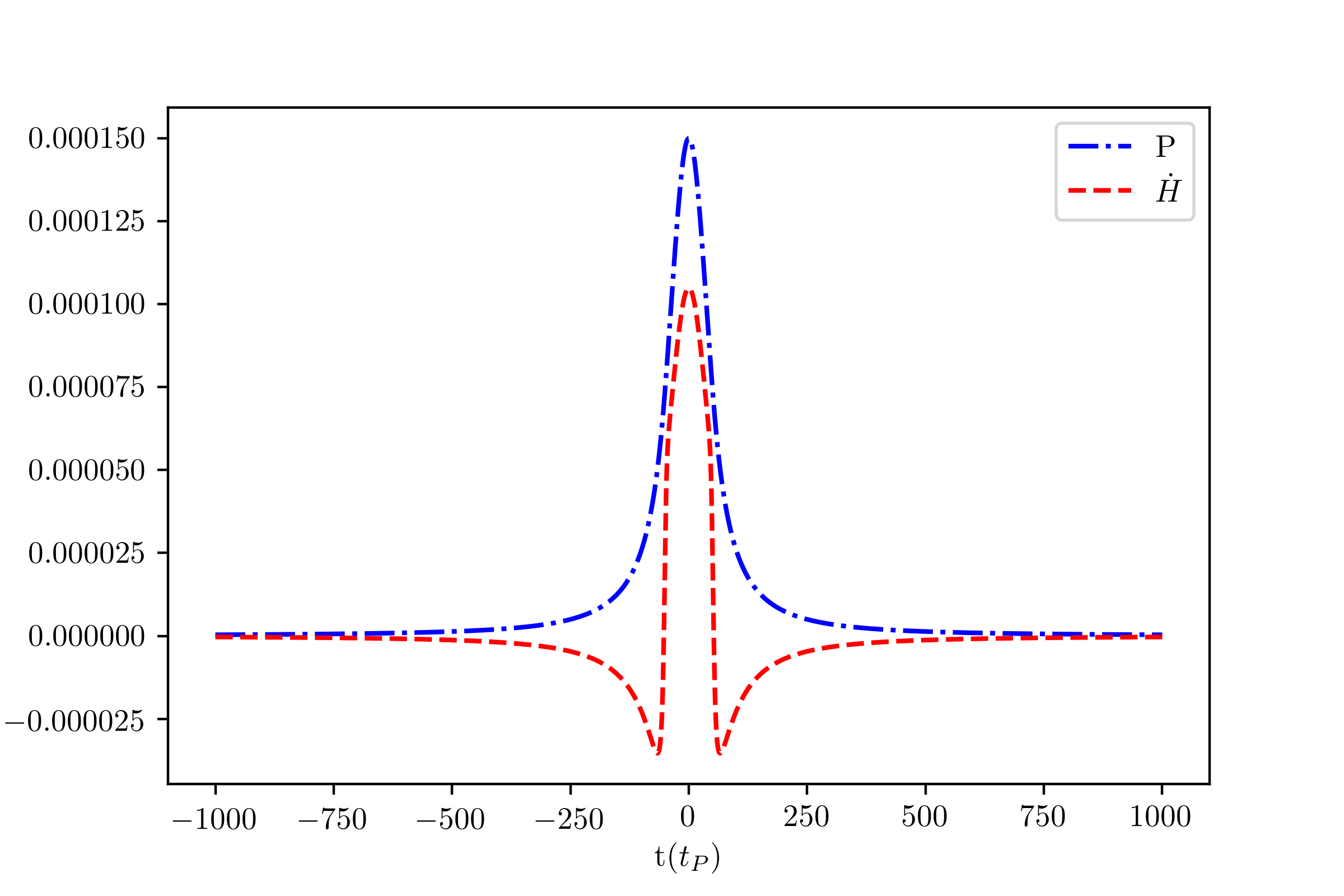}
    \caption{The quantities, $P$, and $\dot{H}$ plotted as a function of time. It can be seen that both quantities are of the same order at the bounce($t=0$)}\label{HdotP}
\end{figure}

As mentioned earlier, the dynamics of the perturbations can be described through the perturbation quantity, $\zeta$. The equations of motion determining the evolution of $\zeta$ was derived in \cite{Cuscuton_paper}. Similar to action, it is convenient to express this equation in the Fourier space

\begin{equation}\label{zeta_eqn}
	\ddot{\zeta}_k + (H+2\frac{\dot{z}}{z})\dot{\zeta}_k+\bigg(\frac{c^2_s k^2}{a^2}\bigg)\zeta_k = 0,
\end{equation}
where the quantities $c_s$ and $z$ are given in equations \eqref{cs} and \eqref{z2} and we find 
\begin{eqnarray}
  2\frac{\dot{z}}{z} &=& -6H-2H\bigg(\frac{((k/a)^2+9P)}{(k/a)^2+3P}\bigg)\nonumber\\ &&+\bigg(\frac{(k/a)^2(2H\dot{H}-2H^3)+P(\ddot{H}-12HP-18H^3)}{(k/a)^2H^2+P(3H^2+P+\dot{H})}\bigg).
\end{eqnarray}

We would like first to point out, that equation \eqref{zeta_eqn} does not become singular  for any value of $k$ at any time. That's because $P>0$,  $P+\dot{H}=\frac{\mu^2}{2M_p^2}|\dot{\varphi}|>0$ and $c_s^2$ is always finite
 \footnote{The denominator of $c^2_s$ is always positive since the quantities $\mathcal{A}_1$ and $\mathcal{A}_2$ simplify to $\mathcal{A}_1 = P(6H^2+\frac{\mu^2}{2}|\dot{\varphi}|)$ and $\mathcal{A}_2= 3P^2(3H^2+\frac{\mu^2}{2}|\dot{\varphi}|)$.}.

We now proceed to numerically explore the dynamics of the perturbations for different scales and as they pass through the bounce. Since at the bounce $H_b=0$ and $\ddot{H}_b=0$, there are two relevant mass scales in equation \eqref{zeta_eqn}, corresponding to quantities $\sqrt{\dot{H}_b}$ and $\sqrt{P}_b$. Figure \ref{HdotP}, demonstrates the time dependence of $\dot{H}$ and $P$ for our model. As we see, both  $\sqrt{\dot{H}_b}$ and $\sqrt{P}_b$ are comparable and around $\sim10^{-2} M_p$ at the bounce. Therefore, we can associate a bounce length scale, $l_B \sim 1/\sqrt{\dot{H}_b}$ to this scale and classify our modes with respect to that. 
We refer to modes as Ultra-Violet (UV)/Infra-Red (IR), if they are shorter/longer with respect to this length scale.

 The equation governing the evolution of the perturbations, \eqref{zeta_eqn} is a second order differential equation, which implies the existence of two independent solutions for each $k$. We have to check the stability for both of these modes to ensure that perturbations are stable on this bouncing background. To do that, we chose two solutions such that one is non-zero at the bounce but has zero derivative there, while the other is zero at the bounce but has non-zero derivative. Since the Wronskian for these solutions is non-zero at the bounce, they are independent.

\begin{figure}[htbb]
    \begin{minipage}{0.49\linewidth}
        \includegraphics[width=\linewidth]{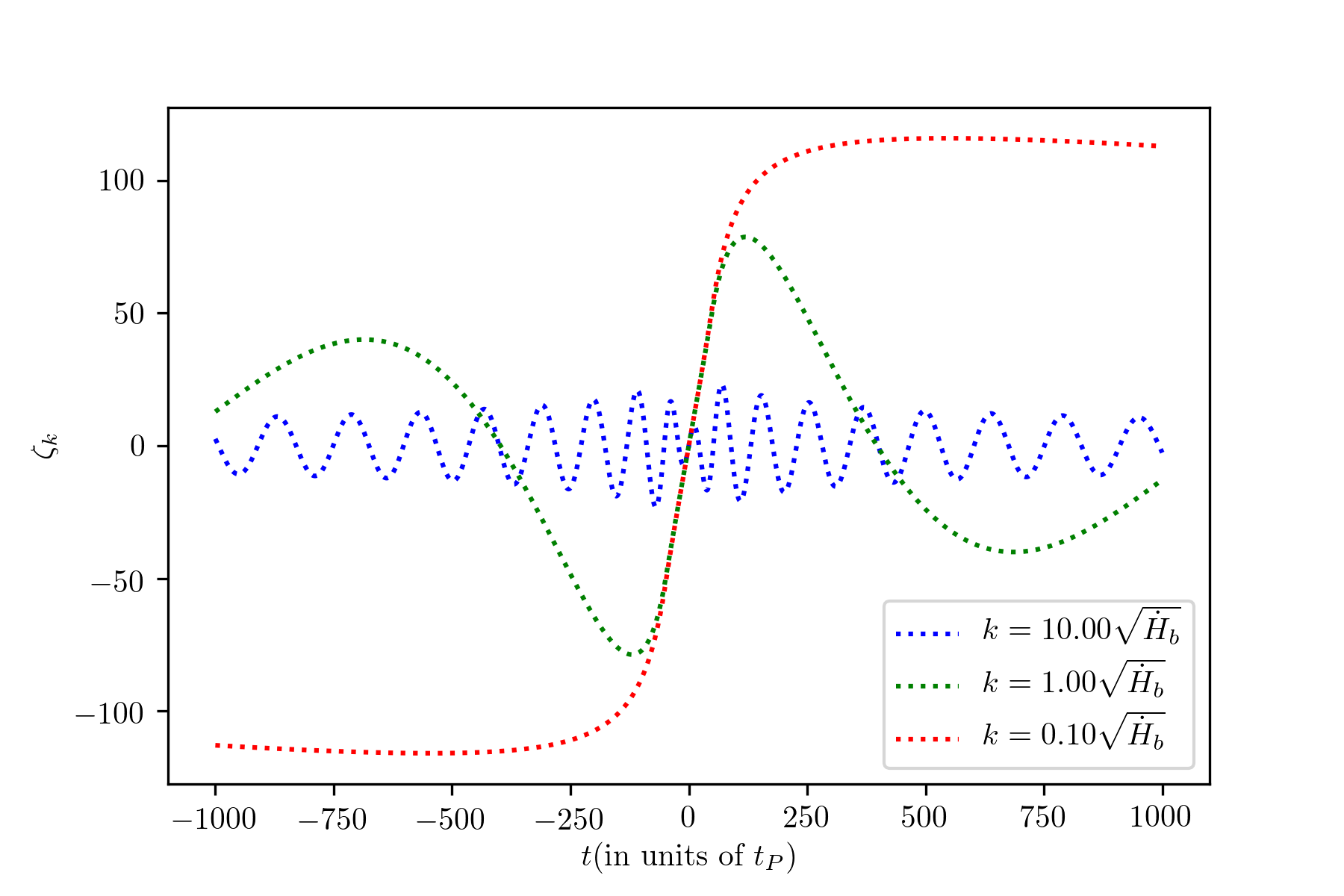}
    \end{minipage}%
    \begin{minipage}{0.49\linewidth}
        \includegraphics[width=\linewidth]{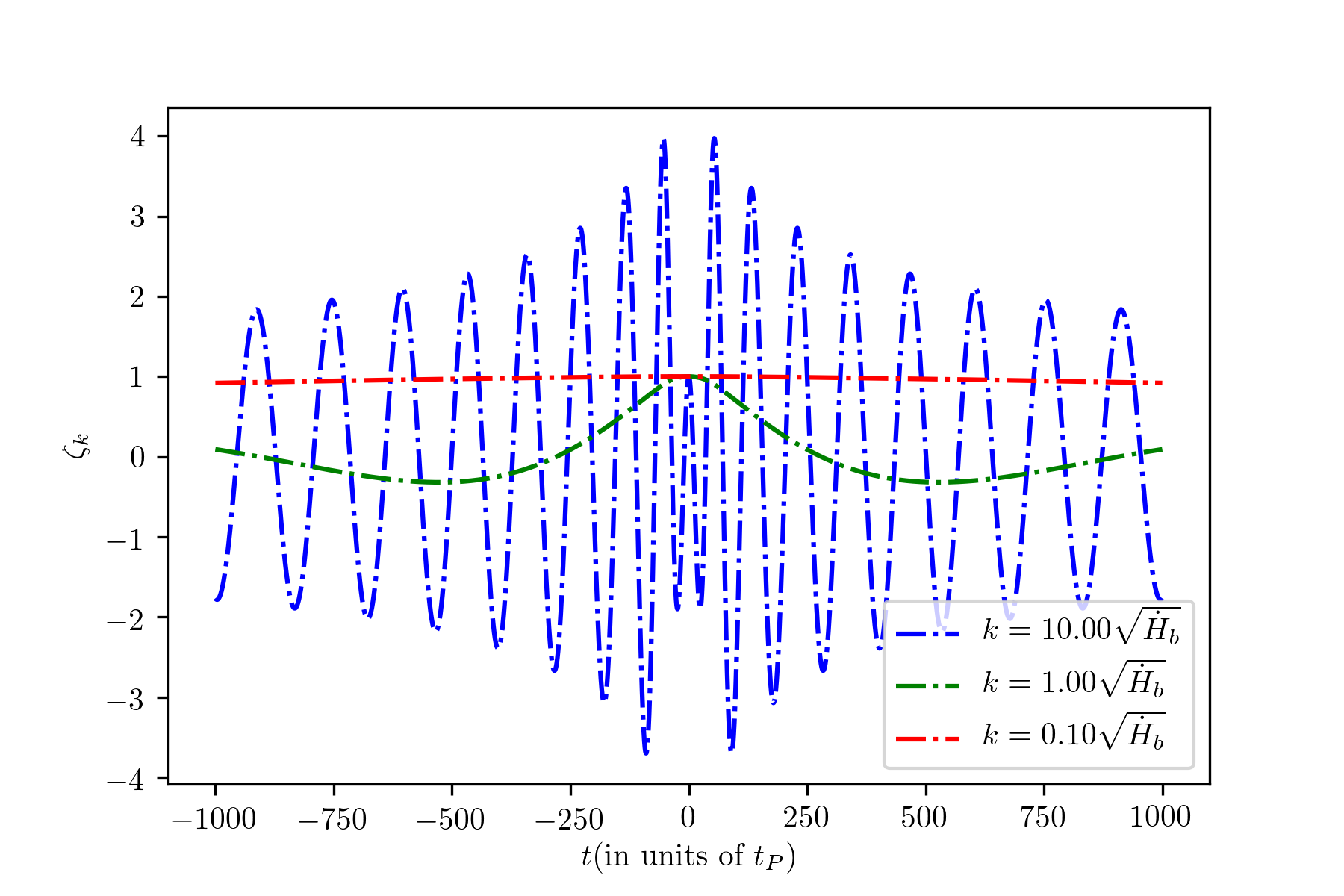}
    \end{minipage}
    \caption{Evolution of perturbations at three different length scales, $k/\sqrt{\dot{H}_b} = 0.1,1.0,10.0$. The two panels correspond to different initial conditions which leads to linearly independent solutions. The left panel has $\zeta_b=0, \dot{\zeta}_b \neq 0 $. The right panel has $\zeta_b \neq 0, \dot{\zeta}_b = 0 $} \label{Pert_evolve}
\end{figure}

To examine the evolution in different regimes, we evolved three wavelength modes, with $\lambda = 0.1l_B, l_B, 10l_B$ numerically.
The results of the numerical evolution for the two independent solutions, is shown in Figures \ref{Pert_evolve}. Our result confirms that there are no instabilities associated with the evolution of modes in different wavelengths scales. As we mentioned before, the value of $\sqrt{\dot{H}_b}$ is $~0.01M_{Pl}$. Therefore, the wavelengths we are investigating are of the order of $10\ell_P, 100\ell_P$ and $1000\ell_P$.   

We conclude that there is no pathology associated with the perturbations at the bounce or at the transition into \cancel{NEC} region ($|t|\sim 60\, t_p$), neither for UV or IR or intermediate scales. 

\section{Conclusions}\label{conclusion}
In this paper, we found a cuscuton bounce solution that has no pathologies associated with \cancel{NEC}. Our solution corresponded to a toy model consisting of a cuscuton field, $\varphi$, in addition to a dynamical matter source, $\pi$. At the background level, we required that away from the bounce (in the contracting or expanding phase) cuscuton density be sub-dominant to matter density. However, we looked for a cuscuton potential such that it would grow faster than matter density as universe contracted and would make the background bounce into expansion. After finding an appropriate potential, we used the cosmological perturbation theory to scrutinize the existence of ghosts and other instabilities in the model. We found that the theory is healthy. We think the underlying reason for absence of instabilities in our model, is that unlike GR, the field which governs the background, i.e. cuscuton, does not have its own dynamical degree of freedom.
Therefore, we expect our result can be extended beyond bounce models to more generic classes of solutions with \cancel{NEC}, which otherwise suffer from instabilities.

\textbf{Acknowledgments} The idea of this project started many years ago by G.G. and over the years it has benefited from discussions with many insightful colleagues. We would like to thank Yifu Cai specially, who made some assessments about this project when it was at its early staged. We also like to thank J. Leo Kim for providing very helpful feedback on our manuscript. This project was supported by the Discovery Grant from Natural Science and Engineering Research Council of Canada (NSERC). G.G. is supported partly by Perimeter Institute (PI) as well. Research at PI is supported by the Government of Canada through the Department of Innovation, Science and Economic Development Canada and by the Province of Ontario through the Ministry of Research, Innovation and Science.

\bibliographystyle{JHEP}
\bibliography{cbounce}




%

\end{document}